\documentclass[12pt]{article}
\usepackage{epsfig}                               

\usepackage{scicite}

\usepackage{times}
\usepackage{amssymb,amsmath}

\newenvironment{sciabstract}{%
\begin{quote} \bf}
{\end{quote}}

\newcounter{lastnote}

\title{Radio Detections During Two State Transitions of the Intermediate Mass Black Hole HLX-1}

\author {Natalie Webb,$^{1,2\ast}$ David Cseh,$^{3}$ Emil Lenc,$^{4,5}$ 
  Olivier Godet,$^{1,2}$  \\ Didier Barret,$^{1,2}$ Stephane
  Corbel,$^{3}$ Sean Farrell,$^{5,6}$ Robert Fender,$^{7}$ \\ Neil
  Gehrels,$^{8}$ Ian
  Heywood$^{9}$\\  \\ \normalsize{$^{1}$Universit\'e de Toulouse;
    UPS-OMP; IRAP, Toulouse, France,}\\
\normalsize{$^{2}$CNRS; IRAP; 9 avenue du Colonel Roche, BP 44346,
  F-31028 Toulouse Cedex 4, France}\\  \normalsize{$^{3}$Laboratoire Astrophysique des
  Interactions Multi-echelles (UMR
  7158),}\\ \normalsize{CEA/DSM-CNRS-Universit\'e Paris Diderot, CEA
  Saclay, F-91191 Gif sur Yvette, France}\\ \normalsize{$^{4}$Australia
  Telescope National Facility, CSIRO Astronomy and Space Science,}
\\ \normalsize{PO Box 76, Epping NSW 1710,
  Australia} \\ \normalsize{$^{5}$Sydney
  Institute for Astronomy, School of Physics, The University of
  Sydney, }\\ \normalsize{NSW 2006, Australia }\\ \normalsize{$^{6}$
  Department of Physics and Astronomy, University of
  Leicester,}\\  \normalsize{University Road, Leicester, LE1 7RH,
  UK}\\ \normalsize{$^{7}$School of Physics and Astronomy, University
  of Southampton, Highfield, }\\ \normalsize{Southampton, SO17 1BJ,
  UK}\\ \normalsize{$^{8}$Astroparticle Physics Laboratory,
  NASA/Goddard Space Flight Center, Greenbelt, MD 20771,
  USA}\\ \normalsize{$^{9}$University of Oxford, Department of
  Physics, Keble Road, Oxford OX1 3RH, UK}\\ \\ \normalsize{$^\ast$To
  whom correspondence should be addressed; E-mail:
  Natalie.Webb@irap.omp.eu} }

\date{}

\begin{document} 


\baselineskip24pt


\maketitle


\begin{sciabstract}

Relativistic jets are streams of plasma moving at appreciable
fractions of the speed of light.  They have been observed from stellar
mass black holes ($\sim$3$-$20 solar masses, M$_\odot$) as well as
supermassive black holes ($\sim$10$^6$$-$10$^9$ M$_\odot$) found in the centres of most galaxies. Jets should also be produced by
intermediate mass black holes ($\sim$10$^2$$-$10$^5$ M$_\odot$), although
evidence for this third class of black hole has until recently been
weak.  We report the detection of transient radio emission
at the location of the intermediate mass black hole candidate ESO
243-49 HLX-1, which is consistent with a discrete jet ejection event. These observations also allow us to refine the mass estimate of the black hole to be between $\sim$9 $\times$10$^{3}$~M$_\odot$ and $\sim$9 $\times$10$^{4}$~M$_\odot$.

\end{sciabstract}



It has been proposed that the kinetic power output of any black
hole is only related to the mass accreted (in Eddington units \cite{EddNote}) on to it and does not depend
on the mass of the black hole itself \cite{hein03}.  If this is true, jet emission,
which is most frequently detected through radio emission, is not only
to be expected from stellar mass black holes 
and supermassive black holes, but
black holes of all masses.  This includes ultra-luminous X-ray sources (ULXs), which are non-nuclear extra-galactic X-ray point sources that exceed the
Eddington luminosity (where the radiation force is balanced by the
gravitational force) for a stellar mass black hole.  These could be either stellar mass black holes undergoing hyper accretion \cite{bege02,glad09} and/or beaming \cite{king01,koer02}. Alternatively they could contain black holes of a slightly higher mass, between 30 and 90 solar masses \cite{zamp09} or be the missing class of intermediate mass black holes \cite{mill04}.  To
date, no variable radio emission associated with jets has been detected from
ultra-luminous X-ray sources, despite numerous observing
campaigns \cite{koer05,free06}.  On the
contrary, non-varying nebula-like extended radio emission, which is
likely to be powered by the central black hole, has been detected around some ULXs \cite{kaar03a,mill05,cseh12}.  ESO 243-49 HLX-1 (hereafter just referred to as HLX-1) is not only a ULX, but currently the best
intermediate mass black hole candidate.  If HLX-1 harbours an intermediate mass black hole, it accretes at comparable fractions of the Eddington luminosity, as stellar mass black holes in binaries. Hence one expects HLX-1 to display similarities with the latter class of objects. In that respect spectral state transitions reminiscent of black hole binaries have already been reported \cite{gode09,serv11}. It is therefore the ideal object in which to search for jet
emission, in order to verify the scale-invariance of jets from black
holes.

HLX-1 was detected serendipitously using XMM-Newton on 23 November
2004 in the outskirts of the edge-on spiral galaxy ESO 243-49, 8" from
the nucleus \cite{farr09}.  The distance to HLX-1 measured
from its H$_\alpha$ emission line confirms that ESO 243-49 is the host
galaxy \cite{wier10}.  HLX-1 therefore has a maximum
unabsorbed X-ray luminosity, assuming isotropic emission, of 1.1
$\times$ 10$^{42}$ ergs s$^{-1}$ \cite{farr09}. The non-nuclear situation of this
point source and the fact that it exceeds the Eddington luminosity
for a stellar mass black hole by three orders of magnitude,
qualify it as a ULX.  From the X-ray
luminosity and the conservative assumption that this value exceeds the
Eddington limit by at most a factor of 10 \cite{bege02}, a lower
limit of 500 M$_\odot$ was derived for the mass of the black hole
\cite{farr09}.  The maximum mass, however, is not constrained.
The X-ray to optical flux ratio \cite{sori10,farr09} is far greater than expected from an AGN, but without an estimate of this maximum
mass, it could be argued that HLX-1 is a non-nuclear supermassive
black hole e.g. \cite{gued09}.

We observed HLX-1 with the  Australia Telescope Compact Array (ATCA) in the 750 m configuration on
13 September 2010 (supporting material \cite{support}), when regular X-ray monitoring of HLX-1 with the Swift satellite \cite{gehr04} showed that HLX-1 had just undergone a transition from the low/hard X-ray state to the high/soft X-ray state.  The transition occurs for HLX-1 when the count rate increases by more than a factor 10 in just a few days (Fig 1) \cite{gode11,serv11}.  Galactic
black hole binaries are known to  regularly emit radio flares
around the transition from the low/hard to the high/soft state,
e.g. \cite{fend09,corb04}.  These are associated with 
ejection events, where, for example, the jet is expelled which can lead to radio flaring when the higher velocity ejecta may collide with the lower-velocity material produced by the steady jet.  As well as detecting radio emission from the nucleus of the galaxy, we
detected a radio point source at Right Ascension (RA) = 01$^h$10$^m$28.28$^s$ and declination (dec.) = -46$^\circ$04'22.3'', coincident with the Chandra X-ray position of HLX-1 \cite{webb10}.  Combining the 5 GHz and 9 GHz data gives a detection of
50~$\mu$Jy/beam, and a 1 $\sigma$ noise level of 11~$\mu$Jy, thus a
4.5 $\sigma$ detection at the position of HLX-1, at a time when such emission can be expected (Fig 2, left) (Table~1).  

The radio flares in Galactic black hole binaries are typically a
factor 10-100 (and even more) brighter than the
non-flaring radio emission \cite{koer05} and generally
last one to several days, e.g. XTE J1859+226 \cite{broc02}.
Once the high/soft state has been achieved, the core jet is suppressed, e.g. \cite{fend09}.  To determine whether  the
radio emission that we detected was transient and thus associated with
a radio flare, we made another observation with the ATCA in the 6 km configuration on 3 December
2010, when HLX-1 was declining from the high/soft state and when no
flaring is expected.  This observation again showed emission from the
nucleus of the galaxy, consistent with that of the previous radio
observation, but revealed no source at the position of HLX-1.   The
3~$\sigma$ non-detection for the combined  5 GHz and 9 GHz data is
36~$\mu$Jy/beam (Fig 2, right) (Table~1). These observations suggest
that the source is variable.  

To confirm the variability, we re-observed HLX-1 when it had just undergone another transition from the low/hard X-ray state to the high/soft X-ray state in August 2011 (Fig 1). All five of the 2011 observations (Table 1) were made in a similar configuration to the December 2010 observation.  We observed three non-contiguous detections ($\geq$ 4 $\sigma$) and two non-contiguous non-detections of the source (Table~1).  This indicates that two flares were detected during this period.

To determine if the source was indeed variable, we fitted each observation using a point source, using the point spread function. We used the position of HLX-1 when the source was not detected.  This allowed us to estimate the flux and the associated errors (Table 1) even for a non-detection. We tested whether the data could be fitted with a constant, namely the mean of the data.  We compared these data to the mean flux value using a chi-squared test.   We found a reduced chi-square ($\chi^{\scriptscriptstyle 2}_{\scriptscriptstyle \nu}$) value of 2.5 (5 degrees of freedom) which is much greater than unity, demonstrating that a constant is a poor fit to  the data and supporting the variable nature. Further, combining all of the detections (5 and 9 GHz), the source is observed at 45~$\mu$Jy/beam, with a 1~$\sigma$ noise level of 5.5~$\mu$Jy, which shows a confident detection at the 8~$\sigma$ level.  Combining, in a similar fashion, the data in which no radio emission was detected, we obtained a  3~$\sigma$ upper limit in the combined 5+9 GHz data of 21~$\mu$Jy/beam (Fig 2).  The variability rules out emission from a nebula.   The observed variable radio emission is then again consistent with a transient jet ejection event.

\begin{table}
{\bf Table 1:} The 7 radio observations organised by date and showing the Swift X-ray unabsorbed flux (0.5$-$10.0 keV) $\times$ 10$^{-13}$
erg cm$^{-2}$ s$^{-1}$ (and the 90\% confidence errors) along with the combined 5 and 9 GHz peak brightness radio flux (with the associated 1 $\sigma$ noise level) or the 3~$\sigma$ upper limit for the non-detections.  The final column gives the radio flux from fitting a point source (using the point spread function) (and the associated 1 $\sigma$ noise level).  \\

\begin{tabular}{cccc}
\hline
Observation & X-ray & 5+9 GHz peak & 5+9 GHz flux \\
date & flux & flux ($\mu$Jy/beam) &  density ($\mu$Jy)\\
\hline
13 Sep. 2010 & 4.57($\pm^{0.68}_{0.50}$) & 50 (11)  & 42 (10)\\
3 Dec. 2010 & 2.40($\pm^{0.60}_{0.50}$) & $<$36 & 11 (20) \\
25 Aug. 2011 & 4.57($\pm$0.30) & $<$30 & 14.5 (7)\\
31 Aug. 2011 & 4.57($\pm$0.30)&  51 (10) & 63 (18)\\
1 Sep. 2011 & 4.57($\pm$0.30)& $<$31 & 25 (10.5) \\
3 Sep. 2011 & 4.57($\pm$0.30)&  45 (10.5) & 43 (10)\\
4 Sep. 2011 & 4.57($\pm$0.30) & 30 (7.5) & 27 (7.5)\\
\hline
\end{tabular}
\end{table}

It has been shown that observations of super massive black holes and stellar mass black holes support the scale invariance of jets \cite{merl03,koer06}.  This was done by comparing X-ray and radio measurements, tracers of
mass accretion rate and kinetic output respectively, with the black hole
mass to form a ``fundamental plane of black hole activity''. Under the hypothesis that HLX-1 is indeed an intermediate mass black hole, we can test the proposed relation.  We take what is generally considered to be the maximum mass of intermediate mass black holes, $\sim$1$\times$10$^{5}$~M$_\odot$ \cite{mill04} and the X-ray luminosity, 5.43$\times$ 10$^{41}$ erg s$^{-1}$ (0.5-10.0 keV), determined from Swift X-ray telescope \cite{burr05} observations made at the same time as our radio detection.  Continuum (non flaring) radio emission could then be estimated with  the aforementioned relationship \cite{koer06},
which is based on a sample that includes black holes in all different X-ray
states.  This relation implies a continuum radio emission at the $\sim$20$\mu$Jy level.  This is slightly lower than the 3~$\sigma$ non-flaring upper limit, suggesting that the mass of the black hole is likely to be less than $\sim$1$\times$10$^{5}$~M$_\odot$.

Radio flares are seen to occur in Galactic black hole binaries when the X-ray luminosity is 10$-$100 per cent of the Eddington luminosity \cite{fend04}.   HLX-1 has already shown similar behaviour to the  Galactic black hole binaries. Therefore assuming that the radio flares that we observed also occur when the X-ray luminosity is 10$-$100 per cent of the Eddington luminosity indicates a black hole mass between $\sim$9.2 $\times$10$^{3}$~M$_\odot$ and $\sim$9.2 $\times$10$^{4}$~M$_\odot$, commensurate with the mass estimate above and those of \cite{davi11,serv11,gode11} and confirming the intermediate mass black hole status.

\bibliography{ATCA2010_rewrite3.bib}

\begin{thebibliography}{10}
\bibitem{EddNote}
The mass accreted (\.m) in Eddington units is \.m = \.M/\.M$_{EDD}$, a
  dimensionless accretion rate which is the accretion rate (\.M) divided by the
  maximum rate of accretion due to the Eddington luminosity (\.M$_{EDD}$).
\end{thebibliography}

\bibliographystyle{Science}

\section*{Acknowledgments}
DC and SC received funding from the European Community's Seventh Framework Programme (FP7/2007-2013) under grant agreement number ITN 215212 "Black Hole Universe".  SF acknowledges funding from the STFC in the UK. SF is the recipient of an Australian Research Council Postdoctoral Fellowship, funded by grant DP110102889.  The Australia Telescope is funded by the Commonwealth of Australia
for operation as a national Facility managed by CSIRO and the data can be accessed at http://atoa.atnf.csiro.au/query.jsp.  This work
made use of data supplied by the UK Swift Science Data Centre
at the University of Leicester (http://www.swift.ac.uk/swift\_portal/)."

\begin{figure}[!h]
\includegraphics[angle=0,scale=0.32]{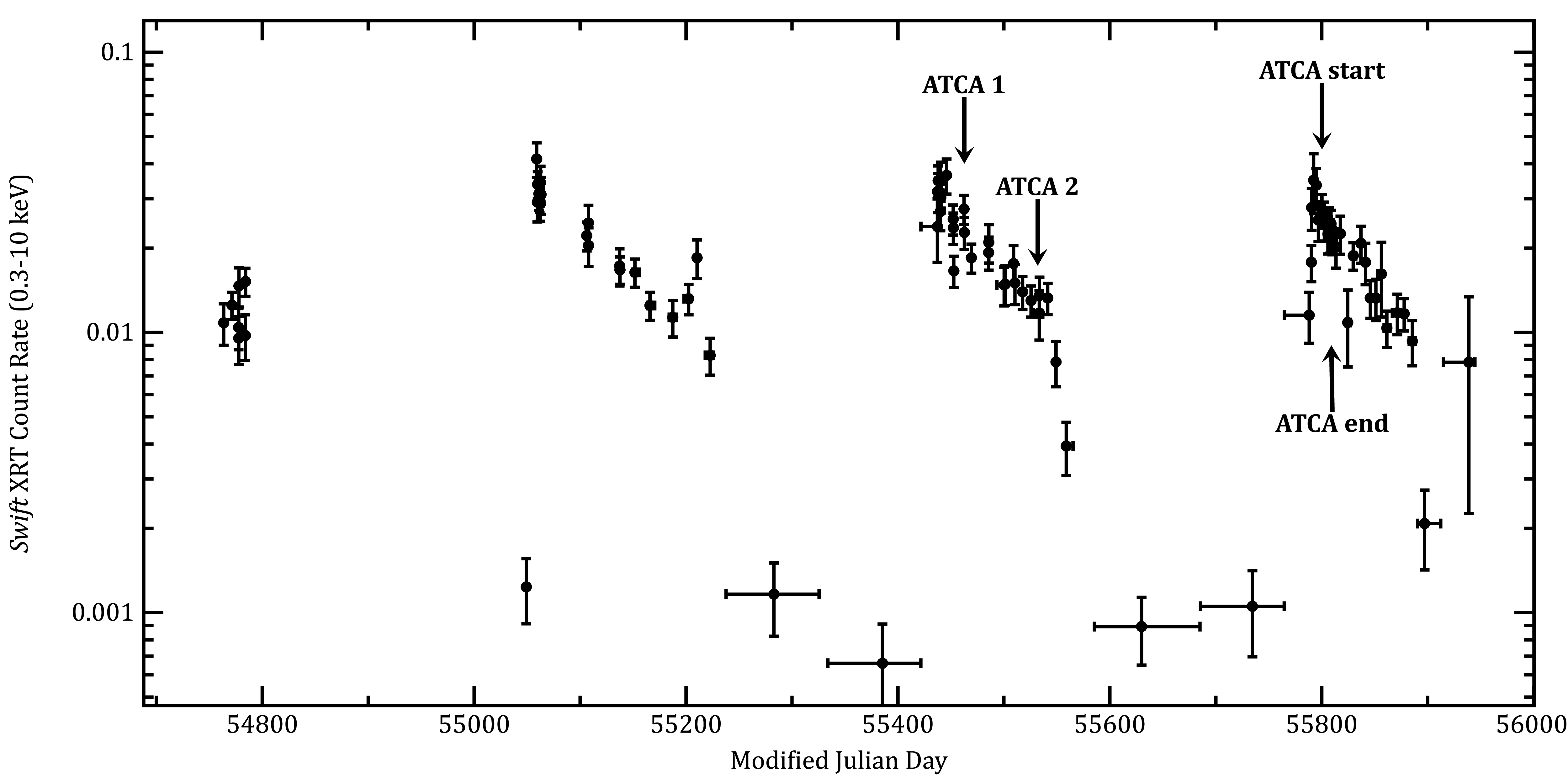}
\caption{Longterm Swift lightcurve showing the dates of the first two ATCA observations and the period during which the subsequent five ATCA observations were taken.  Three X-ray state transitions from the low/hard state (count rate $\lesssim$ 0.002, 0.3-10.0 keV) to the high/soft state (0.01 $\lesssim$ count rate $\lesssim$ 0.05, 0.3-10.0 keV) can be seen.}
\end{figure}

\begin{figure}[!h]
\includegraphics[angle=-90,scale=0.285]{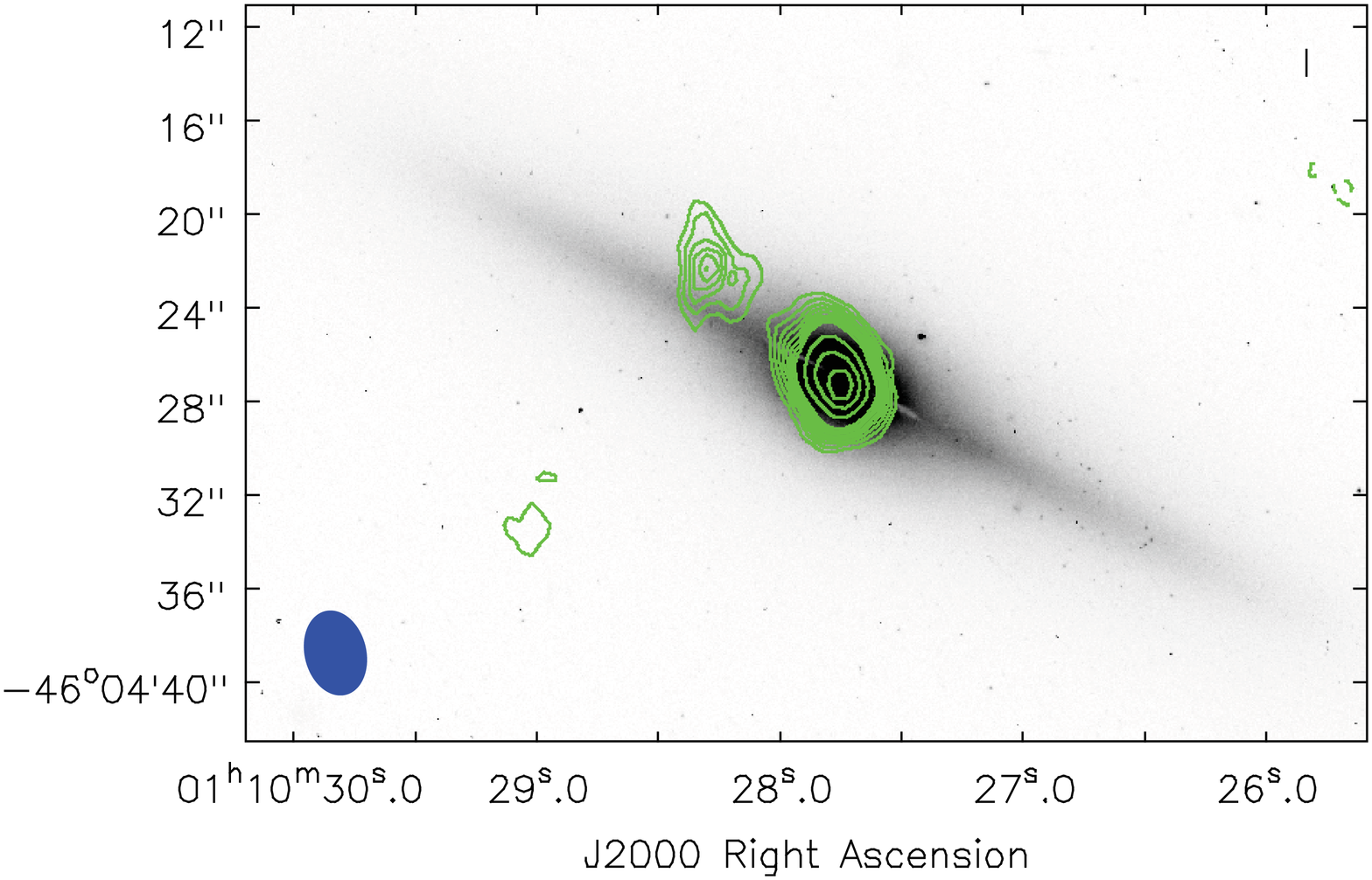}
\includegraphics[angle=-90,scale=0.285]{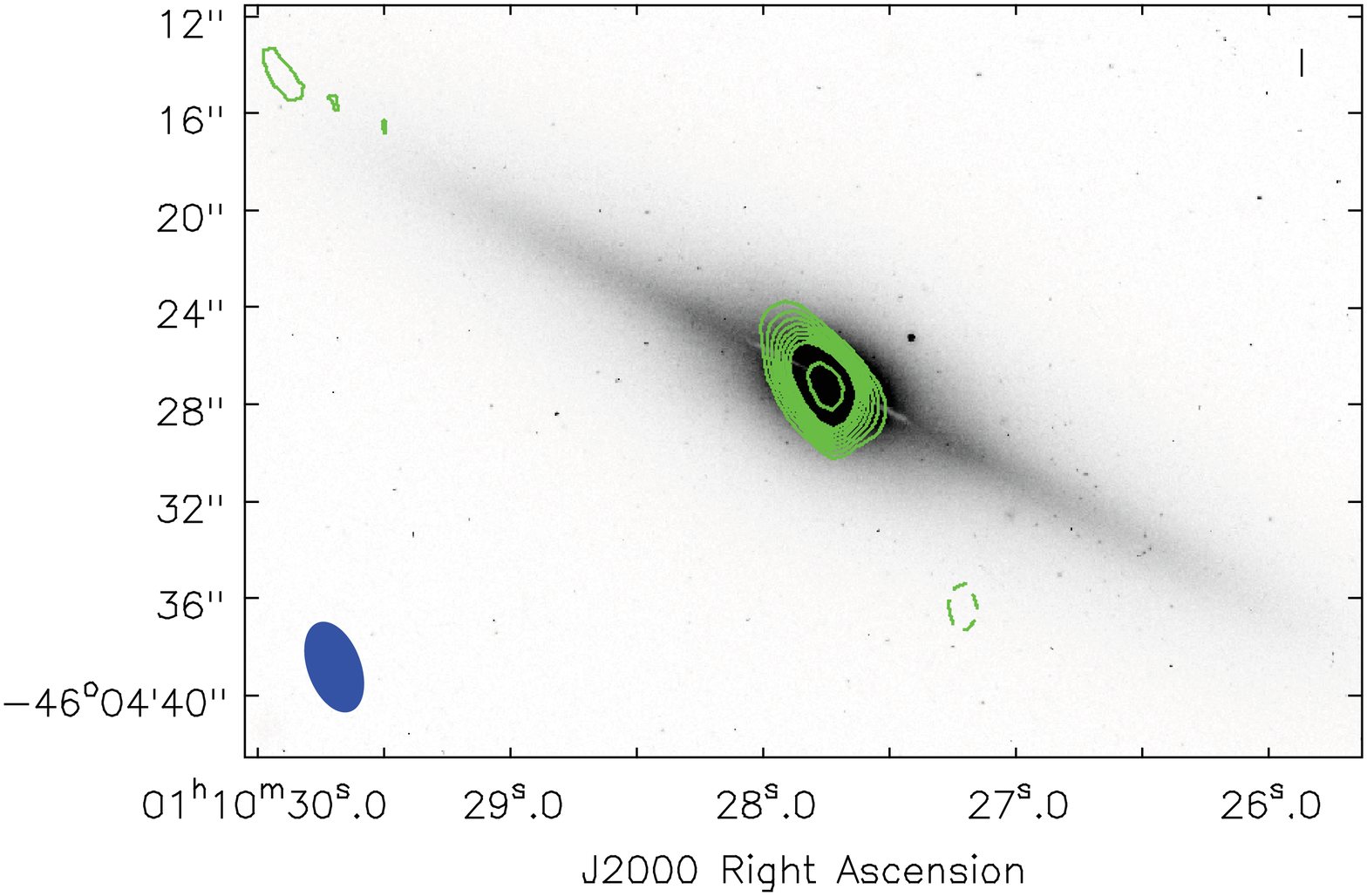}
\caption{Left: 5 and 9 GHz combined radio observations (contours: -3, 3, 4, 5, 6, 7, 8, 9, 10, 15, 20, 25 times the 1 $\sigma$ rms noise level (5.6 $\mu$Jy/beam)) using the radio data taken on the 13th September 2010, 31st August 2011, 3rd and 4th September 2011 with the ATCA and superimposed on an I-band Hubble Space telescope image of ESO 243-49 (inverted colour map). The beam size is shown in the bottom left hand corner.  The galaxy, ESO 243-49, is clearly detected in radio. A 8 $\sigma$ point source falls at RA = 01$^h$10$^m$28.28$^s$ and declination = -46$^\circ$04'22.3'' (1 $\sigma$ error on the position of RA=0.43'' and dec.=0.67''), well within the 0.3'' Chandra error circle of HLX-1. Right: 5 and 9 GHz combined radio observations (contours: -3, 3, 4, 5, 6, 7, 8, 9, 10, 15, 20, 25 times the 1 $\sigma$ rms noise level (7.0 $\mu$Jy/beam)) made from the 3rd December 2010, 25th August 2011 and 1st September 2011 ATCA observations and superimposed on the same I-band Hubble Space telescope image of ESO 243-49.  The galaxy ESO 243-49 is again clearly detected, but no source is found within the Chandra error circle.  Again the beam size is shown in the bottom left hand corner.   }
\end{figure}

\section*{Supporting Material}

\section*{Observations and analysis}
HLX-1 in ESO 243-49 was observed with the Australia Telescope Compact Array seven times as shown in Table~S1, using the upgraded Compact Array Broadband Backend (CABB) \cite{wils11}. The data were taken using the CFB 1M-0.5k correlator configuration with 2 GHz bandwidth and 2048 channels, each with 1 MHz resolution. Each observation was performed at the central frequencies of 5.5 GHz and 9 GHz simultaneously. During the first observation the array was in the 750 m configuration (giving baselines up to 5 km when all 6 antennas are used).  During the second and subsequent observations it was in the 6 km configuration. The total on-source integration time was $\sim$11 h for each observation. The primary calibrator PKS 1934-638 was used for absolute flux and bandpass calibration, while the secondary calibrator 0048-427 was used for the phase and antenna gain calibration. For each observation, we observed 1934-638 for 10 min and the phase calibrator was observed every 15 min. 

The data reduction and analysis was performed with the Multichannel Image Reconstruction, Image Analysis and Display (MIRIAD) software \cite{saul95}. We loaded the data into MIRIAD using the {\sc atlod} task with options birdie, xycorr, rfiflag, and noauto, which flags out the channels affected by self-interference, correcting the phase difference between the X and Y channels, discarding any autocorrelation data, and automatically flagging out frequency bands that are known to be heavily affected by Radio Frequency Interference (RFI). The standard data reduction steps were flagging, bandpass, phase and amplitude calibration, following the MIRIAD User Guide\footnote{http://www.atnf.csiro.au/computing/software/miriad/userguide/userhtml.html}. We used multi-frequency synthesis (MFS) methods \cite{wils11} to produce the dirty maps. Imaging was carried out using the multi-frequency \cite{wils11} clean algorithms. We note that imaging did not involve any self-calibration.

The detections with the associated 1 $\sigma$ noise level and the 3~$\sigma$ non-detections for the 5 GHz, 9 GHz and combined observations are given in Table~S1. All the images were naturally weighted in order to reach the best sensitivity, with the exception of the September 2010 5 GHz (and the 5+9 GHz) image which was weighted with robust=0 due to the lower resolution of this data. We also combined the 5-GHz and 9-GHz data sets in order to enhance sensitivity. 

We estimated the position errors by adding the errors due to phase calibration, the position of the phase calibrator and the point-source model fit in quadrature. We note that as we have a Very Long Baseline Interferometry measurement for the phase calibrator, its positional error is negligible. The error due to phase calibration depends on the distance between the target and the calibrator of 11.5 degrees which corresponds to a positional uncertainty of $\sim$0.2''. We find the error on the RA is 0.43'' and on the dec. is 0.67''.       

We verified that the radio source was consistent with a point-like object by comparing the fluxes and the fitted point source model fluxes.  For all of our detections, the radio source was consistent with a point source.

\begin{table}
{\bf Table S1:} The 7 radio observations organised by date and showing the radio detection with the associated 1 $\sigma$ noise level or the 3~$\sigma$ non-detection for the 5 GHz, 9 GHz and combined observations\\

\begin{tabular}{ccccccc}
\hline
 & \multicolumn{2}{c}{5 GHz flux ($\mu$Jy/beam)} & \multicolumn{2}{c}{9 GHz flux ($\mu$Jy/beam)} & \multicolumn{2}{c}{5+9 GHz flux ($\mu$Jy/beam)} \\
\hline
Observation & Detection & Non-detect. & Detection  &  Non-detect. & Detection  & Non-detect. \\
date & (1~$\sigma$ noise) & (3~$\sigma$) & (1~$\sigma$ noise) &  (3~$\sigma$) & (1~$\sigma$ noise) & (3~$\sigma$)\\
\hline
13 Sep. 2010 & 45 (11) &  & & 36 & 50 (11) & \\
3 Dec. 2010 &  & 33 & & 63 & & 36\\
25 Aug. 2011 & & 36 & & 45 & & 30 \\
31 Aug. 2011 & 44 (11.5) & &  & 51 &  51 (10) & \\
1 Sep. 2011 & & 36 & & 48 & & 31 \\
3 Sep. 2011 & & 36 &  81 (17) & & 45 (10.5) & \\
4 Sep. 2011 & & 27 & & 39 & 30 (7.5) & \\
\hline
\end{tabular}
\end{table}

The Swift-XRT Photon
Counting data were processed using the tool XRTPIPELINEv0.12.6. We
used the grade 0-12 events, giving slightly higher effective area at
higher energies than the grade 0 events, and a 20 pixel radius circle
to extract the source and background spectra using XSELECT v2.4b.  The
ancillary response files were created using XRTMKARF v0.5.9 and
exposure maps generated by XRTEXPOMAP v0.2.7. We fitted all the
spectra within XSPEC v12.7.0 using the response file
SWXPC0TO12S6-20070901V012.RMF \cite{gode09b}. 

The 11 ks of data taken on the 13-14th September 2010 revealed a spectrum that was well fitted with an absorbed
disc blackbody model ($\chi^{\scriptscriptstyle 2}_{\scriptscriptstyle
  \nu}$=0.66, 7 degrees of freedom (dof), n$_H$ = 4 $\times$ 10$^{20}$ cm$^{2}$,
kT=0.20$^{\scriptscriptstyle  +0.02}_{\scriptscriptstyle -0.02}$ keV).
This gives a 0.5-10.0 keV unabsorbed flux of  4.57($\pm^{\scriptscriptstyle 0.68}_{\scriptscriptstyle 0.50}$) $\times$ 10$^{-13}$
erg cm$^{-2}$ s$^{-1}$, thus a luminosity of 4.9 $\times$ 10$^{41}$ erg
s$^{-1}$ (0.5-10.0 keV), assuming a source distance of 95 Mpc (using the
WMAP cosmology).  

The 6.4 ks of Swift data taken on the 3rd December 2010 gave a spectrum that we fitted with the same blackbody and 
absorbed power law model as the second X-ray observation presented in \cite{farr09}.  These characteristics were chosen as the faintness of the source meant that there were too few counts to constrain the fit.  The goodness of the fit was acceptable  (C-stat=38, 61 channels) and the unabsorbed  flux value was
found to be 2.4$\pm^{\scriptscriptstyle 0.60}_{\scriptscriptstyle 0.50}$) $\times$ 10$^{-13}$ erg cm$^{-2}$ s$^{-1}$ (0.5-10.0 keV) which
gives  an unabsorbed luminosity of 2.7$\times$ 10$^{41}$ erg s$^{-1}$
(0.5-10.0 keV).  

Data taken during the 2011 outburst were best fitted with the same model as the September 2010 data.  We initially took all the data during the observing period from 5th August 2011 to the 16th August 2011 (34 ks of data) and found an unabsorbed flux value of 4.57$\pm0.30 \times$ 10$^{-13}$ erg cm$^{-2}$ s$^{-1}$ (0.5-10.0 keV). Using a subset of this data (taken between the 24th and 31st August 2011, which gives 7.2 ks X-ray data) and fitting with the same model reveals an unabsorbed flux of 4.86$\pm0.70 \times$ 10$^{-13}$ erg cm$^{-2}$ s$^{-1}$ (0.5-10.0 keV), hence compatible within the 90\% confidence errors quoted.  We therefore took the flux calculated over the whole period, as it has the smallest error bars, which reveals a luminosity of 5.43$\times$ 10$^{41}$ erg s$^{-1}$ (0.5-10.0 keV).

\section*{The $\chi^2$ test to determine the variability of the source}

We tested the hypothesis that the flux values are constant.  Fitting the seven flux values determined through the point source fitting (Table 1), in association with the errors, with a constant value we determined a mean flux of 28.48 $\mu$Jy.  The 1 $\sigma$ errors on this value are $\pm$3.71$\mu$Jy. The $\chi^2$ of the fit was found to be 12.52.  Since the mean value was obtained from the sample data, the number of degrees of freedom are the number of data points minus 1 minus the one fitted parameter, which totals five. The  $\chi^{\scriptscriptstyle 2}_{\scriptscriptstyle \nu}$ value of 2.5 is much greater than unity, indicating that a constant is a poor fit to  the data.

\section*{The black hole fundamental plane}

We used the sample presented by \cite{merl03}, which includes black holes in all states and the correlation presented from  the analysis of this sample by \cite{koer06}, as they consider the emission within the 0.5-10.0 keV band, as opposed to the emission in the 2.0-10.0 keV band in the study of \cite{merl03}.  The wider band used by \cite{koer06} is better adapted to the current study of HLX-1, as the X-ray emission in the high/soft state shows very few counts in the 2.0-10.0 keV band, leading to large uncertainties in the X-ray luminosity in this band.  The correlation found by  \cite{koer06} is 

$$ log(L_x) = \xi_R log(L_R) + \xi_M log(M) + b_x $$

where $L_x$ is the X-ray luminosity (0.5-10.0 keV), $\xi_R$ = 1.74$\pm$0.20, $L_R$ is the radio luminosity, $\xi_M$ = -1.35$\pm$0.27, $M$ is the mass of the black hole, $b_x$ = -14.23$\pm$5.75.

\end{document}